\documentclass{article}
\usepackage[english]{babel}
\usepackage{amsmath,amssymb,graphicx,enumerate,bbm,xcolor,theorem}
\usepackage{geometry}
\catcode`\<=\active \def<{
\fontencoding{T1}\selectfont\symbol{60}\fontencoding{\encodingdefault}}
\catcode`\>=\active \def>{
\fontencoding{T1}\selectfont\symbol{62}\fontencoding{\encodingdefault}}
\catcode`\|=\active \def|{
\fontencoding{T1}\selectfont\symbol{124}\fontencoding{\encodingdefault}}
\newcommand{\assign}{:=}
\newcommand{\nobracket}{}
\newcommand{\nocomma}{}
\newcommand{\nosymbol}{}

\newcommand{\tmdummy}{$\mbox{}$}
\newcommand{\tmem}[1]{{\em #1\/}}
\newcommand{\tmop}[1]{\ensuremath{\operatorname{#1}}}
\newcommand{\um}{-}
\newenvironment{enumeratealpha}{\begin{enumerate}[a{\textup{)}}] }{\end{enumerate}}
\newtheorem{corollary}{Corollary}
\newtheorem{definition}{Definition}
{\theorembodyfont{\rmfamily}\newtheorem{example}{Example}}
\newtheorem{proposition}{Proposition}
{\theorembodyfont{\rmfamily}\newtheorem{question}{Question}}
{\theorembodyfont{\rmfamily}\newtheorem{remark}{Remark}}
\begin{document}

\title{Measure Concentration on Fermi Balls}
\author{Kurt Pagani \\
 \small{\emph{Dedicated to Maex on her 55th birthday}}
}
\date{June 26,\, 2014}

\maketitle

\begin{abstract}
  This note reports on some attempts to examine if and under which conditions
  the naturally scaled probability measures associated to an orthonormal basis
  of a classical Paley-Wiener space converge to a uniform distribution (on a compact set in
  momentum space). The results are still quite unsatisfactory, yet we got some
  indications that for inf-compact functions (symbols) and domains for which
  some generalized Weyl law holds, the measures converge weakly to a
  (generalized) Fermi ball. \ \ \ 
\end{abstract}

\section{Notation and Results}

Let $\Omega$ be an open subset of $\mathbbm{R}^{n}$ with finite Lebesgue
measure, i.e. $\mathcal{L}^{n} ( \Omega ) < \infty .$ The classical
Paley-Wiener space $P W_{\Omega} ( \mathbbm{R}^{n} ) = \left\{ f \in C (
\mathbbm{R}^{n} ) \cap L^{2} ( \mathbbm{R}^{n} ) :  \, \tmop{supp} ( \check{f}
)   \subset   \bar{\Omega} \right\}$ is a reproducing kernel Hilbert space
with kernel $Q_{\Omega} ( \xi , \eta ) = ( 2 \pi )^{-n/2}  
\hat{\chi}_{\Omega} ( \xi - \eta )$, where $\chi_{\Omega}$ denotes the
characteristic function of $\Omega$ and $\hat{f} , \check{f}$ are the
(unitary) Fourier and inverse Fourier transform of $f$ respectively. The space
$L^{2} ( \Omega )$ is (unitarily) isomorphic to $P W_{\Omega} (
\mathbbm{R}^{n} )$ if we extend the elements of $L^{2} ( \Omega )$ by zero
outside $\mathbbm{R}^{n} \backslash \Omega \nobracket .$ The well known
Paley-Wiener states that $P W_{\Omega} ( \mathbbm{R}^{n} )$ is the set of
restrictions of entire functions of exponential type to $\mathbbm{R}^{n}$,
whence these functions are real analytic and the set $P W_{\Omega} (
\mathbbm{R}^{n} ) \cap \{ \| f \| =1 \}$ is uniformly bounded, i.e. $\| f
\|_{\infty} \leqslant \frac{| \Omega |}{( 2 \pi )^{n}} .$ The $L^{p}  $norm on
$\mathbbm{R}^{n}$ is denoted by $\| \cdot \|_{p}$ where we usually omit the
subscript when $p=2.$

In the following let $\Phi :\mathbbm{R}^{n} \rightarrow [ 0, \infty ]$ be an
inf-compact function, that is the level sets $\{ \xi : \Phi ( \xi ) \leqslant
t \}$ are compact (or empty) for all $t \geqslant 0.$ We fix an orthonormal
basis $\{ \varphi_{m} \}_{m \in \mathbbm{N}}$ of $P W_{\Omega} (
\mathbbm{R}^{n} )$ and define a ``spectrum'' of $\Phi$ as follows:
\begin{equation}
  \lambda_{m} ( \Phi ) = \int_{\mathbbm{R}^{n}} \Phi ( \xi )   | \varphi_{m} (
  \xi ) |^{2}  d \xi , \hspace{1.2em} m=1,2,3 \ldots \label{eq1}
\end{equation}
It is clear that the $\lambda_{m}$ may be infinite and depend on the basis
chosen, however, to simplify notation we do not indicate this dependence but
one should keep this fact in mind when looking for minimizers. In the same
spirit we associate to a basis $\{ \varphi_{m} \}_{m \in \mathbbm{N}}$ two
sets of probability measures $\{ \nu_{N} \}_{N \in \mathbbm{N}}   \subset 
\mathcal{P} ( \mathbbm{R}^{n} )$ and $\{ \mu_{N} \}_{N \in \mathbbm{N}}$
$\subset  \mathcal{P} ( \mathbbm{R}^{n} ) :$
\begin{equation}
  \nu_{N} ( f ) = \frac{1}{N} \sum_{m=1}^{N} \int_{\mathbbm{R}^{n}} f ( \xi ) 
  | \varphi_{m} ( \xi ) |^{2}  d \xi , \hspace{1.2em} N=1,2,3, \ldots
  \label{eq2}
\end{equation}
and
\begin{equation}
  \mu_{N} ( f ) = \sum_{m=1}^{N} \int_{\mathbbm{R}^{n}} f ( \xi )   |
  \varphi_{m} ( N^{1/n}   \xi ) |^{2}  d \xi , \hspace{1.2em} N=1,2,3, \ldots
  \label{eq3}
\end{equation}
for all $f \in C_{0} ( \mathbbm{R}^{n} )$ = continuous functions with compact
support. It is easily seen that $S_{N,\#} \mu_{N} = \nu_{N}$ where $\#$
denotes push-forward in this context and $S_{N}$ is the scaling map $( S_{N} f
) ( \xi ) =f ( N^{1/n} \xi ) .$

\begin{definition}
  Given $\Phi$ and $\{ \varphi_{m} \}_{m \in \mathbbm{N}}$ as above we set \
  $M ( \Lambda ) = \{ m \in \mathbbm{N}:  \lambda_{m} ( \Phi ) \leqslant
  \Lambda \}$ and denote by $\mathcal{N} ( \Lambda )$ the number of elements
  in $M ( \Lambda ) .$\label{def1}
\end{definition}

We have the following upper bound:

\begin{proposition}
  \label{p2}{\tmdummy}
  
  \begin{equation}
    {\color{black} \mathcal{N} ( \Lambda ) \leqslant \frac{| \Omega |}{( 2 \pi
    )^{n}} \inf_{0< \varepsilon <1}   \frac{\mathcal{L}^{n} \left( \left\{
    \Phi \leqslant \tfrac{\Lambda}{\varepsilon} \right\} \right)}{1-
    \varepsilon}} \label{eq4}
  \end{equation}
\end{proposition}

For any orthonormal basis $\{ \varphi_{m} \}_{m \in \mathbbm{N}}$ of $P
W_{\Omega} ( \mathbbm{R}^{n} )$ the kernel $Q_{\Omega}$ has the representation
\[ Q_{\Omega} ( \xi , \eta ) = ( 2 \pi )^{-n/2}   \hat{\chi}_{\Omega} ( \xi -
   \eta ) = \sum_{m=1}^{\infty} \varphi_{m} ( \xi )   \bar{\varphi}_{m} ( \eta
   ) \]
and the trace is bounded on the whole of $\mathbbm{R}^{n}$, indeed
\begin{equation}
  \lim_{N \rightarrow \infty}   \sum_{m=1}^{N} | \varphi_{m} ( \xi ) |^{2} =
  \frac{| \Omega |}{( 2 \pi )^{n}} \label{eq5}
\end{equation}
uniformly on compact subsets (also follows from Bessel's inequality).

\begin{definition}
  The Fermi ball associated to $\Omega$ is determined by the requirement
  \[ \frac{| \Omega |}{( 2 \pi )^{n}} | B_{\kappa_{F}} | =1. \]
  Thus the radius $\kappa_{F}$ is given by
  \begin{equation}
    \kappa_{F} = \frac{2 \pi}{( \omega_{n} | \Omega | )^{1/n}} , \label{eq6}
  \end{equation}
  where $\omega_{n}$ is the volume of the unit ball in $\mathbbm{R}^{n}
  .$\label{def3}
\end{definition}

\begin{example}
  Set $\Phi_{p} ( \xi ) = | \xi |^{p}$, $p>0$, then $\Phi$ is certainly
  inf-compact and we get
  \[ \mathcal{N} ( \Lambda ) \leqslant \frac{| \Omega |}{( 2 \pi )^{n}}
     \inf_{0< \varepsilon <1}   \frac{\omega_{n}   \left(
     \tfrac{\Lambda}{\varepsilon} \right)^{n/p}}{1- \varepsilon} = \frac{|
     \Omega |   \omega_{n}   \Lambda^{n/p}}{( 2 \pi )^{n}}   \inf_{0<
     \varepsilon <1}   \frac{\varepsilon^{-n/p}}{1- \varepsilon} =C_{n,p }
     \frac{| \Omega |   \omega_{n}   \Lambda^{n/p}}{( 2 \pi )^{n}} . \]
  The function $\frac{\varepsilon^{-n/p}}{1- \varepsilon}$ attains its minimum
  in $( 0,1 )$ at $\varepsilon_{0} = \frac{n}{n+p}$ so that $C_{n,p} =
  \frac{n+p}{p  \left( \frac{n}{n+p} \right)^{n/p}} .$ Now if $N=2 
  \tilde{C}_{n}   \left( \frac{\Lambda^{1/p}}{\kappa_{F}} \right)^{n}
  \geqslant 2N ( \Lambda )$ then we have (at least half of the $\lambda_{m}$
  must be >$  \Lambda$):
  \[ \sum_{j=1}^{N} \lambda_{j} ( \Phi_{p} ) \geqslant \frac{N}{2}   \Lambda
     \geqslant   \frac{N}{2}   \left( \frac{N}{2 \widetilde{C_{n}}}
     \right)^{p/n} \kappa_{F}^{p}   \sim  N^{1+p/n}   \kappa_{F}^{p} . \]
  \[  \]
  where $\tilde{C}_{n} \geqslant C_{n,p}$ such that $N$ is an integer.
\end{example}

\begin{definition}
  For any N=1,2,3,.. we define the quantities
  \[ \omega_{\Phi} ( N ) := \frac{\int_{\mathbbm{R}^{n}} \Phi ( N^{-1/n}   \xi
     )  d \nu_{N} ( \xi )}{N  \int_{\mathbbm{R}^{n}} \Phi (   \xi )  d \nu_{N}
     ( \xi )} . \]
  For instance if $\Phi$ is homogeneous of degree $p$ then $\omega_{\Phi} ( N
  ) =N^{- \left( 1+ \frac{p}{n} \right)} .$\label{def5}
\end{definition}

\begin{proposition}
  Let $\Phi$ and $\{ \varphi_{m} \}_{m \in \mathbbm{N}}$ as above and suppose
  that
  \begin{equation}
    \sup_{N}   \left[ \omega_{N} ( \Phi )   \sum_{m=1}^{N} \lambda_{m} ( \Phi
    ) \right] < \infty \label{supc}
  \end{equation}
  then the sequence of measures $\{ \mu_{N} \}_{N \in \mathbbm{N}} \subset
  \mathcal{P} ( \mathbbm{R}^{n} )$ is tight and therefore (by Prokhorov's
  theorem) a subsequence $\mu_{N_{j}}$ converges weakly to a measure
  $\mu_{\star} \in \mathcal{P} ( \mathbbm{R}^{n} )$. Moreover we have that
  \[ \liminf_{j \rightarrow \infty}   \int_{\mathbbm{R}^{n}} \Phi ( \xi )  d
     \mu_{N_{j}} ( \xi ) \geqslant   \int_{\mathbbm{R}^{n}} \Phi ( \xi )  d
     \mu_{\star} ( \xi ) . \]
\end{proposition}

It might be instructive to consider the case $\Phi = \Phi_{p} .$\label{p6}

\begin{example}
  If $\Phi ( \xi ) = | \xi |^{p}$ then we get
  \[ \int_{\mathbbm{R}^{n}} | \xi |^{p}  d \mu_{N} ( \xi ) =N^{- \left( 1+
     \frac{p}{n} \right)}   \sum_{m=1}^{N} \lambda_{m} \]
  which, when setting $p=2$ and choosing the basis $\{ \varphi_{m} \}_{m} = \{
  \hat{u}_{m} \}_{m}$, where $\Delta u_{m} + \lambda_{m} u_{m} =0$ in $\Omega$
  and $u_{m} =0$ in $\mathbbm{R}^{n} \backslash \Omega \nobracket$, yields by
  the well known Weyl law (provided that it is valid for $\Omega$):
  \[ \lim_{N \rightarrow \infty}  N^{- \left( 1+ \frac{2}{n} \right)}  
     \sum_{m=1}^{N} \lambda_{m} = \frac{n}{n+2}   \kappa_{F}^{2} . \]
  In a similiar way we get when using Weyl's law for the fractional Laplacian
  [G]:
  \[ \lim_{N \rightarrow \infty}  N^{- \left( 1+ \frac{p}{n} \right)}  
     \sum_{m=1}^{N} \lambda_{m} = \frac{n}{n+p}   \kappa_{F}^{p} . \]
  It is obvious that the right hand side above is $\int_{B_{\kappa_{F}}} | \xi
  |^{p}  d  \xi$. Since ($\ref{supc}$) is certainly satisfied one may wonder
  if $d \mu_{\star} = \chi_{B_{\kappa_{F}}} ( \xi )  d \xi$ holds.
\end{example}

A variant of the ``bathtub'' lemma yields:

\begin{proposition}
  \label{p8}Let $\Phi$ and $\{ \varphi_{m} \}_{m \in \mathbbm{N}}$ be given,
  then for all $N \in \mathbbm{N}:$

  \[ \int \Phi ( \xi )  d \mu_{N} ( \xi )   \geqslant   \frac{| \Omega |}{( 2
     \pi )^{n}} \int_{\{ \Phi < \tau \}} \Phi ( \xi )  d \xi +c_{0}    \frac{|
     \Omega | \tau}{( 2 \pi )^{n}} \mathcal{L}^{n} ( \{ \xi : \Phi ( \xi ) =
     \tau \} ) , \]
  where
  \begin{equation}
    \tau = \sup \left\{ t:\mathcal{L}^{n} ( \{ \Phi <t \} ) \leqslant \frac{(
    2 \pi )^{n}}{| \Omega |} \right\} \label{eq7}
  \end{equation}
  and if $\mathcal{L}^{n} ( \{ \xi : \Phi ( \xi ) = \tau \} ) >0  $ then
  \[ c_{0} = \frac{\frac{( 2 \pi )^{n}}{| \Omega |} -\mathcal{L}^{n} ( \{ \xi
     : \Phi ( \xi ) < \tau \} )}{\mathcal{L}^{n} ( \{ \xi : \Phi ( \xi ) =
     \tau \} )}  . \label{eq8} \]
  Otherwise $c_{0} =0.$
\end{proposition}

Now it is apparent to suspect that for a ``minimal'' basis of $P W_{\Omega} (
\mathbbm{R}^{n} )$ the measures $\mu_{N}$ converge vaguely (at least) or even
better, weakly to
\begin{equation}
  d \mu_{\Phi} = \frac{| \Omega |}{( 2 \pi )^{n}} [   \chi_{\{ \Phi < \tau \}}
  ( \xi ) +c_{0}   \chi_{\{ \Phi = \tau \}} ( \xi ) ]  d \xi . \label{eq9}
\end{equation}
For the example $\Phi = | \xi |^{p}$ again we get $\tau = \kappa_{F}^{p}$ and
we can choose $c_{0} =0$ since $\mathcal{L}^{n} ( \{ \xi : | \xi | =
\kappa_{F} \} ) =0.  $

\begin{corollary}
  Let $\Phi$ and $\{ \varphi_{m} \}_{m \in \mathbbm{N}}$ be given and assume
  that
  \[ \Lambda ( \Phi ) \assign \liminf_{N \rightarrow \infty}   \omega_{N} (
     \Phi )   \sum_{m=1}^{N} \lambda_{m} ( \Phi ) = \int_{\mathbbm{R}^{n}}
     \Phi  d \mu_{\Phi}  < \infty \]
  where $d \mu_{\Phi} = \frac{| \Omega |}{( 2 \pi )^{n}} [   \chi_{\{ \Phi <
  \tau \}}   ( \xi ) +c_{0}   \chi_{\{ \Phi = \tau \}} ( \xi ) ]  d \xi$, then
  there is a subsequence $\{ N_{j} \}_{}$ and a measure $\mu_{\star} \in
  \mathcal{P} ( \mathbbm{R}^{n} )$ such that $\mu_{N_{j}} ( f ) \rightarrow  
  \mu_{\star} ( f )$,$\forall f \in C_{b} ( \mathbbm{R}^{n} )$ and
  \[ \lim_{j \rightarrow \infty}   \int_{\mathbbm{R}^{n}} \Phi ( \xi )  d
     \mu_{N_{j}} ( \xi ) = \int \Phi ( \xi )  d \mu_{\star} ( \xi ) = \Lambda
     ( \Phi ) . \]
  Moreover the basis $\{ \varphi_{m} \}_{m \in \mathbbm{N}}$ is minimal in the
  sense that $\Lambda ( \Phi )$ is the smallest value than can be achieved by
  the associated sequence of measures $\{ \mu_{N}^{\varphi} \}_{N \geqslant 1}
  ._{}$\label{c9}
\end{corollary}

\begin{question}
  When do we have $\mu_{\star} = \mu_{\Phi}$ ?\label{q10}
\end{question}

Obviously one is inclined trying to prove that $d \mu_{\star} ( \xi )$=
$\varphi ( \xi )  d \xi$, for some $\varphi \in L^{\infty} ( \mathbbm{R}^{n}
)$, and that $\| \varphi \|_{\infty} \leqslant \frac{| \Omega |}{( 2 \pi
)^{n}}$. Then we had that $\varphi \in \mathcal{C}$ and consequently $\varphi$
would be a minimizer. Then a condition like $\mathcal{L}^{n} ( \{ \xi : \Phi (
\xi ) = \tau \} ) =0$ guarantees uniqueness and we were done. However, due to
the non-separability of $L^{\infty} ( \mathbbm{R}^{n} )$, it is by no means
granted that the sequence
\[ G_{N} ( \xi ) = \sum_{m=1}^{N} | \varphi_{m} ( N^{1/n} \xi ) |^{2} \]
has a (weak$\star$) convergent subsequence in $L^{\infty} ( \mathbbm{R}^{n} )
.$ Yet we can show:

\begin{proposition}
  Under the conditions of Corollary $\ref{c9}$, the limit measure is
  absolutely continuous with respect to Lebesgue measure: $d \mu_{\star} ( \xi
  ) =G_{\star} ( \xi )  d \xi$, with $G_{\star} \in L^{p} ( \mathbbm{R}^{n} )$
  for all $1 \leqslant p \leqslant \infty$ and
  \[ \| G_{\star} \|_{p} \leqslant \left( \frac{| \Omega |}{( 2 \pi )^{n}}
     \right)^{1-1/p} , \forall p \; \tmop{such} \; \tmop{that}  1 \leqslant p<
     \infty . \]
  Moreover we have $\| G_{\star} \|_{1} =1.$ $\label{p11}$
\end{proposition}

Now, since we know that $G_{\star} \in L^{\infty} ( \mathbbm{R}^{n} )$ we also
have that $\| G_{\star} \|_{\infty} \leqslant \frac{| \Omega |}{( 2 \pi
)^{n}}$ by taking the limit $p \rightarrow \infty$ above. Therefore we get a
partial answer to question $\left( \ref{q10} \right) :$

\begin{corollary}
  Under the conditions of Corollary $\ref{c9}$, and if moreover
  $\mathcal{L}^{n} ( \{ \xi : \Phi ( \xi ) = \tau \} ) =0$, then
  \[ d \mu_{\star} = \frac{| \Omega |}{( 2 \pi )^{n}}   \chi_{\{ \Phi < \tau
     \}}   ( \xi )  d \xi , \]
  that is $G_{\star} ( \xi ) = \frac{| \Omega |}{( 2 \pi )^{n}}   \chi_{\{
  \Phi < \tau \}}   ( \xi ) .$
\end{corollary}

\section{Proofs}

\subsection{Proposition $\ref{p2}$}

By definition $\ref{def1}$ we have for $m \in M ( \Lambda )$ and $t>0:$
\[ t \int_{\{ \Phi >t \}} | \varphi_{m} ( \xi ) |^{2}   \leqslant \int_{\{
   \Phi >t \}} \Phi ( \xi ) | \varphi_{m} ( \xi ) |^{2}   \leqslant
   \lambda_{m} ( \Phi ) = \int_{\mathbbm{R}^{n}} \Phi ( \xi ) | \varphi_{m} (
   \xi ) |^{2}   \leqslant \Lambda , \]
thus
\[ \int_{\{ \Phi \leqslant t \}} | \varphi_{m} ( \xi ) |^{2} \geqslant 1-
   \frac{\Lambda}{t} , \forall m \in M ( \Lambda ) ,t>0. \]
Therefore by $\left( \ref{eq5} \right) :$
\[ \mathcal{N} ( \Lambda )   \left( 1- \frac{\Lambda}{t} \right) \leqslant
   \sum_{m \in M ( \Lambda )} \int_{\{ \Phi \leqslant t \}} | \varphi_{m} (
   \xi ) |^{2}  d \xi \leqslant \frac{| \Omega |}{( 2 \pi )^{n}} 
   \mathcal{L}^{n} ( \{ \Phi \leqslant t \} ) . \]
Set $\Lambda = \varepsilon  t$, then we get
\[ \mathcal{N} ( \Lambda ) ( 1- \varepsilon ) \leqslant \frac{| \Omega |}{( 2
   \pi )^{n}}  \mathcal{L}^{n} \left( \left\{ \Phi \leqslant
   \frac{\Lambda}{\varepsilon} \right\} \right) \]
for any $\varepsilon \in ( 0,1 )$ which proves $\left( \ref{eq4} \right) .$

\subsection{Proposition $\ref{p6}$}

By assumption $C= \sup_{N}   \left[ \omega_{N} ( \Phi )   \sum_{m=1}^{N}
\lambda_{m} ( \Phi ) \right] < \infty$. Therefore we have for all $N \geqslant
1:$
\[ \frac{\int_{\mathbbm{R}^{n}} \Phi ( N^{-1/n}   \xi )  d \nu_{N} ( \xi )}{N 
   \int_{\mathbbm{R}^{n}} \Phi (   \xi )  d \nu_{N} ( \xi )}   \sum_{m=1}^{N}
   \lambda_{m} ( \Phi ) \leqslant C . \]
By the relations $\ref{eq1} , \ref{eq2}$ and $\ref{eq3}$ we get
\[ \int_{\mathbbm{R}^{n}} \Phi ( \xi )  d \mu_{N} ( \xi ) = \frac{1}{N}
   \int_{\mathbbm{R}^{n}} \Phi ( N^{-1/n}   \xi )  d \nu_{N} ( \xi ) \leqslant
   C. \]
Recall that a subset $A$ of $\mathcal{P} ( \mathbbm{R}^{n} )$ is called
{\tmem{tight}} if for any $\varepsilon >0$ exists a compact subset
$K_{\varepsilon}$ of $\mathbbm{R}^{n}$ such that $\mu ( K_{\varepsilon} )
\geqslant 1- \varepsilon$ for all $\mu \in A.$ When we set $K_{j} =L_{\Phi} (
j ) = \{ \xi :  \Phi ( ) \leqslant j \} ,j \in \mathbbm{N},$then for any $\mu
\in \{ \mu_{m} \}_{m \in \mathbbm{N}}$
\[ j  \mu ( \mathbbm{R}^{n} \backslash \nobracket K_{j} ) \leqslant  
   \int_{\mathbbm{R}^{n} \backslash K_{j} \nobracket} \Phi ( \xi )  d \mu (
   \xi )   \leqslant \int_{\mathbbm{R}^{n}} \Phi ( \xi )  d \mu_{} ( \xi )
   \leqslant  C, \]
thus
\[ 1- \mu ( K_{j} ) \leqslant \frac{C}{j}   \Rightarrow   \mu ( K_{j} )
   \geqslant 1- \frac{C}{j} . \]
Since $\Phi$ is inf-compact, all the sets $K_{j}$ are compact by definition,
so this proves that the sequence $\{ \mu_{m} \}_{m \in \mathbbm{N}}$ is tight
in $\mathcal{P} ( \mathbbm{R}^{n} ) .$ Next, by Prokhorov's theorem \cite{PY}, there is
a subsequence $\mu_{m_{j}}$ which converges weakly to a measure $\mu_{\star}
\in \mathcal{P} ( \mathbbm{R}^{n} ) .$ Furthermore we have by the lower
semicontinuity of the function $\Phi$ (which trivially follows by the
compactness of its level sets):
\[ \liminf_{j \rightarrow \infty} \int \Phi ( x )  d \mu_{m_{j}} ( x )
   \geqslant   \int \Phi ( x )  d \mu_{\star} ( x ) . \]

\subsection{Proposition $\ref{p8}$}

Let $\mu \in \mathcal{M} ( \mathbbm{R}^{n} )$ and $\Phi :\mathbbm{R}^{n}
\rightarrow \mathbbm{R}$ such that $\mu ( \{ x: \Phi ( x ) >t \} ) < \infty$
for all $t \in \mathbbm{R}.$ Then we want to show that
\[ \mathcal{E}_{\Phi} ( \varphi_{0} ) = \inf_{\varphi \in \mathcal{C}}   \int
   \Phi ( x )   \varphi ( x )  d \mu ( x ) = \int_{\{ \Phi < \tau \}} \Phi ( x
   )  d \mu ( x ) +c_{0}   \tau   \mu ( \{ x: \Phi ( x ) = \tau \} ) , \]
where
\[ \mathcal{C}= \left\{ \varphi  : 0 \leqslant \varphi ( x ) \leqslant 1, \int
   \varphi d \mu =A \right\} \cap \{ \varphi   \tmop{is}   \mu -
   \tmop{measurable} \} \]
and
\[ \tau = \sup \{ t: \mu ( x: \Phi ( x ) <t ) \leqslant A \} \]
and
\[ c_{0}   \mu ( \{ x: \Phi ( x ) = \tau \} ) =A-  \mu ( x: \Phi ( x ) < \tau
   ) . \]
The minimizer
\[ \varphi_{0} ( x ) = \chi_{\{ \Phi < \tau \}} ( x ) +c \chi_{\{ \Phi = \tau
   \}} ( x ) \]
is unique if either $A= \mu ( x: \Phi ( x ) < \tau )$ or $A= \mu ( x: \Phi ( x
) \leqslant \tau ) .$

To prove it we show that $\mathcal{E}_{\Phi} ( \varphi_{0} ) \leqslant
\mathcal{E}_{\Phi} ( \varphi_{} ) :$ $\tmop{note}   \tmop{that}   \varphi_{0}
=1  \tmop{on}   \{ \Phi < \tau \}$, then (using $[ \ldots ]$ as delimiters):
\[ \int \Phi ( x )   [ \varphi_{0} ( x ) - \varphi ( x ) ]  d \mu ( x ) =  \]
\[ \int_{\{ \Phi < \tau \}} \Phi [ \varphi_{0} - \varphi ] d \mu + \int_{\{
   \Phi > \tau \}} \Phi [ \varphi_{0} - \varphi ] d \mu + \int_{\{ \Phi = \tau
   \}} \Phi [ \varphi_{0} - \varphi ] d \mu \leqslant \]
\[ \tau \int_{\{ \Phi < \tau \}} [ \varphi_{0} - \varphi ]_{} d \mu - \int_{\{
   \Phi > \tau \}} \Phi \varphi d \mu + \tau \int_{\{ \Phi = \tau \}} [
   \varphi_{0} - \varphi ] d \mu \]
\[ \leqslant \tau \int_{\{ \Phi < \tau \}} [ \varphi_{0} - \varphi ]_{} d \mu
   - \tau \int_{\{ \Phi > \tau \}} \varphi d \mu + \tau \int_{\{ \Phi = \tau
   \}} [ \varphi_{0} - \varphi ] d \mu \]
\[ = \tau   \int_{\{ \Phi \leqslant \tau \}} \varphi_{0}  d \mu - \tau \int
   \varphi  d \mu =0. \]

If $\mu ( \{ \Phi < \tau \} ) <A \nocomma ,$ while $\mu ( \{ \Phi = \tau \} )
>A,$ then the difference mass $A- \mu ( \{ \Phi < \tau \} )$ has to be
distributed on the level set $\{ \Phi = \tau \}$, what can be done in several
ways, thus the minimizer is not unique in this case. \

Recalling the form of the measures $\mu_{N}$, see ($\ref{eq3}$), and letting
$A^{-1} = \frac{| \Omega |}{( 2 \pi )^{n}}$ we notice that the functions
$\psi_{m,N} ( x ) =A^{} | \varphi_{m} ( N^{1/n}   \xi ) |^{2}$ are uniformly
bounded by $1$ and therefore elements of the admissible set $\mathcal{C}$, so
that
\[ \frac{| \Omega |}{( 2 \pi )^{n}} \int_{\{ \Phi < \tau \}} \Phi ( \xi )  d
   \xi +c_{0}    \frac{| \Omega | \tau}{( 2 \pi )^{n}} \mathcal{L}^{n} ( \{
   \xi : \Phi ( \xi ) = \tau \} ) \]
is a lower bound to $\mu_{N} ( \Phi )$ at any rate, as has been claimed.

\subsection{Proposition \ref{p11}}

Let $F_{N} ( \xi ) = \sum_{m=1}^{N}   | \varphi_{m} ( \xi ) |^{2}$ and $G_{N}
( \xi ) =F_{N} ( N^{1/n} \xi ) \nosymbol .$ Then the sequences $\{ F_{N} \}_{N
\geqslant 1}$ and $\{ G_{N} \}_{N \geqslant 1}  $are uniformly bounded on
$\mathbbm{R}^{n}$ by $\frac{| \Omega |}{( 2 \pi )^{n}} .$ Now we have
\[ \int_{\mathbbm{R}^{n}} F_{N} ( N^{1/n}   \xi )^{p}  d \xi = \frac{1}{N}  
   \int_{\mathbbm{R}^{n}} F_{N} (   \xi )^{p}  d \xi = \frac{1}{N}  
   \int_{\mathbbm{R}^{n}} F_{N} ( \xi ) F_{N} (   \xi )^{p-1}  d \xi \leqslant
   \left( \frac{| \Omega |}{( 2 \pi )^{n}} \right)^{p-1} , \]
thus $G_{N} \in L^{p} ( \mathbbm{R}^{n} )$ for all $p \geqslant 1,$ and $\|
G_{N} \|_{p} \leqslant \left( \frac{| \Omega |}{( 2 \pi )^{n}} \right)^{1-1/p}
.$ Now, denote by $G_{N_{m}}  $the subsequence which converges to the limit
measure $\mu_{\star}$, that is
\[ \lim_{m \rightarrow \infty} \int \varphi ( \xi )  G_{N_{m}} ( \xi )  d \xi
   = \int \varphi ( \xi )  d \mu_{\star} ( \xi ) = \mu_{\star} ( \varphi ) ,
   \forall \varphi \in C_{b} ( \mathbbm{R}^{n} ) , \]
then
\[ \lim_{m \rightarrow \infty} \int G_{N_{m}} ( \xi )  d \xi = \lim_{m
   \rightarrow \infty}   \| G_{N_{m}} \|_{1} = \mu_{\star} ( 1 ) =1. \]
The set $S \assign \{ G_{N} :N \in \mathbbm{N} \}_{} \subset L^{1} (
\mathbbm{R}^{n} )$ is uniformly integrable. Indeed,
\[ \int_{A} G_{N} ( \xi ) d \xi = \int_{A} F_{N} ( N^{1/n} \xi )  d \xi =
   \frac{1}{N} \int_{N^{1/n} A} F_{N} ( \eta )  d \eta \leqslant \frac{|
   \Omega |}{( 2 \pi )^{n}} | A | . \]
Hence, by the Dunford-Pettis theorem, $S$ is relatively weakly compact in
$L^{1} ( \mathbbm{R}^{n} )$ and thus the Eberlein-Smulian theorem guarantees
the relatively weak sequential compactness of $S.$ So, there is another
subsequence, $G_{N_{m_{j}}}$ which converges weakly in $L^{1} (
\mathbbm{R}^{n} )$ to a function $G_{\star}$, that is
\[ \lim_{j \rightarrow \infty} \int G_{N_{m_{j}}} ( \xi ) f ( \xi )  d \xi =
   \int f ( \xi )  G_{\star} ( \xi )  d \xi , \forall f \in L^{\infty} (
   \mathbbm{R}^{n} ) . \]
Since $C_{b} ( \mathbbm{R}^{n} ) \subset L^{\infty} ( \mathbbm{R}^{n} )$
(although not dense) we must have $d \mu_{\star} ( \xi ) =G_{\star} ( \xi )  d
\xi$, and consequently
\[ \| G_{\star} \|_{1} =1. \]
Clearly, we also have that
\[ G_{N_{m_{j}}} \xrightarrow[\tmop{weak} -L^{p}]{} G_{\star} \]
for all $1<p< \infty$ and therefore
\[ \| G_{\star} \|_{p} \leqslant \liminf_{j \rightarrow \infty}   \left\|
   G_{N_{m_{j}}} \right\|_{p} \leqslant \left( \frac{| \Omega |}{( 2 \pi
   )^{n}} \right)^{1-1/p} , \forall  1 \leqslant p< \infty . \]
It remains the case $p= \infty .$ Suppose $G_{\star} \notin L^{\infty} (
\mathbbm{R}^{n} )$ then we can find a constant $C> \frac{| \Omega |}{( 2 \pi
)^{n}}$ and a measurable set $A \subset \mathbbm{R}^{n}$ satisfying $\infty
>\mathcal{L}^{n} ( A ) >0$ such that $| G_{\star} ( \xi ) | \geqslant C$ for
all $\xi \in A.$ However, this implies
\[ \liminf_{p \rightarrow \infty}   \| G_{\star} \|_{p}   \geqslant  C > 
   \frac{| \Omega |}{( 2 \pi )^{n}} , \]
which contradicts $\| G_{\star} \|_{p} \leqslant \left( \frac{| \Omega |}{( 2
\pi )^{n}} \right)^{1-1/p}$.

\section{Remarks and Examples}
This section provides some examples and thoughts which motivated the investigation reported in the first section. 

\subsection{Dirichlet Laplacian}

Let $\Omega$ be an open subset of $\mathbbm{R}^{n}$ with compact closure
$\bar{\Omega}$ and denote by $\lambda_{j}$, $u_{j} ( x )$, $j \in \mathbbm{N}$
the eigenvalues and eigenfunctions to the Dirichlet problem:
\begin{equation}
  \left\{\begin{array}{l}
    \Delta u ( x ) + \lambda u ( x )  = 0 , \; x \in \Omega\\
    u ( x ) =0  \nocomma , \; x \in \mathbbm{R}^{n} \backslash \Omega
    \nobracket .
  \end{array}\right.   \label{eq10}
\end{equation}
We assume that the eigenfunctions are orthogonal, normalized and ordered by
increasing eigenvalues, thus
\[ \| u_{j} \| =1 \nocomma , \; \| \nabla u_{j} \| = \sqrt{\lambda_{j}} , \]
where $\| f \| = \sqrt{\langle f,f \rangle}$ denotes the $L^{2}$ norm on
$\mathbbm{R}^{n} .$ We use the Fourier transform in the unitary form
\[ \hat{u} ( k ) = ( 2 \pi )^{-n/2} \int_{\mathbbm{R}^{n}} u ( x )  e^{- i 
   \langle k,x \rangle} d x.  \]
Bessel's inequality immediatley shows that the Fourier transforms of the
eigenfunctions to $\left( \ref{eq10} \right)$ are uniformly bounded on
$\mathbbm{R}^{n}$ and moreover:
\[ \lim_{N \rightarrow \infty} \sum_{j=1}^{N}   | \hat{u}_{j} ( k ) |^{2}  =
   \frac{| \Omega |}{( 2 \pi )^{n}} \]
for all $k \in \mathbbm{R}^{n}$, where the convergence is uniform on compact
sets. Since the eigenfunctions are certainly in $L^{1} ( \mathbbm{R}^{n} )$ we
get by the Riemann-Lebesgue lemma that the $\hat{u}_{j}$ are continuous and
\[ \lim_{| k | \rightarrow \infty}   | \hat{u}_{j} ( k ) | =0. \]
Therefore the convergence of the partial sums above cannot be uniform on the
whole of $\mathbbm{R}^{n}$. Furthermore, we have by Parsevals theorem
\[ \sum_{j=1}^{N} \int_{\mathbbm{R}^{n}}   | \hat{u}_{j} ( k ) |^{2}  d k=N \]
and
\[ \sum_{j=1}^{N} \int_{\mathbbm{R}^{n}} | k |^{2}   | \hat{u}_{j} ( k )
   |^{2}  d k=  \sum_{j=1}^{N} \int_{\mathbbm{R}^{n}}   | \nabla u_{j} ( x )
   |^{2}  d x=  \sum_{j=1}^{N} \lambda_{j} , \]
so that both expressions above tend to infinity if $N$ does. If we scale each
sum by the transformation $k \rightarrow  a_{N}  k$, $a_{N} >0,$ we get the
equations:
\[ \sum_{j=1}^{N} \int_{\mathbbm{R}^{n}}   | \hat{u}_{j} ( a_{N}  k ) |^{2}  d
   k= \frac{N}{a_{N}^{n}} \]
and
\[ \sum_{j=1}^{N} \int_{\mathbbm{R}^{n}} | k |^{2}   | \hat{u}_{j} ( a_{N}  k
   ) |^{2}  d k= \frac{1}{a_{N}^{n+2}} \sum_{j=1}^{N} \lambda_{j} , \]
yielding - when eliminating the $a_{N}$ on the right hand sides:
\[ \frac{\sum_{j=1}^{N} \int_{\mathbbm{R}^{n}} | k |^{2}   | \hat{u}_{j} (
   a_{N}  k ) |^{2}  d k}{\left( \sum_{j=1}^{N} \int_{\mathbbm{R}^{n}}   |
   \hat{u}_{j} ( a_{N}  k ) |^{2}  d k \right)^{1+2/n}} =N^{1- \tfrac{2}{n}}  
   \sum_{j=1}^{N} \lambda_{j} . \]
The most obvious choice seems to be $a_{j} =N^{1/n}$ for all $j$. Indeed, the
nominator becomes unity and the last equation simplifies to:
\begin{equation}
  \sum_{j=1}^{N} \int_{\mathbbm{R}^{n}} | k |^{2}   | \hat{u}_{j} ( N^{1/n}  k
  ) |^{2}  d k=N^{1- \tfrac{2}{n}}   \sum_{j=1}^{N} \lambda_{j} . \label{W2}
\end{equation}
We also notice by Fatou's Lemma
\[   \int_{\mathbbm{R}^{n}} \liminf_{N \rightarrow \infty}   \sum_{j=1}^{N}  
   | \hat{u}_{j} ( a_{N}  k ) |^{2}  d k \leqslant   \liminf_{N \rightarrow
   \infty}   \frac{N}{a_{N}^{n}} , \]
that means if we scale too strong then the partial sums may converge to zero
a.e. whenever
\[ \liminf_{N \rightarrow \infty}   \frac{N}{a_{N}^{n}} =0. \]
However, $\lim_{N \rightarrow \infty}   \sum_{j=1}^{N}   | \hat{u}_{j} ( 0 )
|^{2} = \frac{| \Omega |}{( 2 \pi )^{n}}$ exists in any case.

\begin{question}
  Does the limit
  \[ \lim_{N \rightarrow \infty}   \sum_{j=1}^{N}   | \hat{u}_{j} ( N^{1/n} 
     k ) |^{2}   \]
  exists pointwise? 
\end{question}

Apparently, the same procedure can also be repeated for other operators
instead of the Laplacian. If we take the fractional Laplacian $( p>0 )$
instead we obtain
\begin{equation}
  \sum_{j=1}^{N} \int_{\mathbbm{R}^{n}} | k |^{p}   | \hat{u}_{j} ( N^{1/n}  k
  ) |^{2}  d k=N^{1- \tfrac{p}{n}}   \sum_{j=1}^{N} \lambda_{j} , \label{Wp}
\end{equation}
where the meaning of the involved quantities has to be changed
correspondingly, of course. However, the question remains unchanged,
indicating a general phenomenon. It will be instructive to have a look at the
case $n=1$.

\subsubsection{$\tmop{Case}  n=1$}

Let $\bar{\Omega} = [ 0, \pi ]$. Then the eigenfunctions to $\left( \ref{eq1}
\right)$ have the simple form
\[ u_{m} ( x ) = \sqrt{\frac{2}{\pi}}   \chi_{[ 0, \pi ]} ( x )   \sin ( m x )
\]
with corresponding eigenvalues $\lambda_{m} =m^{2} .$ The Fourier transforms
are
\[ \hat{u}_{m} ( k ) = \frac{1}{\pi} \int_{0}^{\pi} \sin ( m x )  e^{- i k x} 
   d x. \]
Now let us calculate
\[ F_{N} ( k ) = \sum_{m=1}^{N} | \hat{u}_{m} ( k ) |^{2} . \]
Instead of using the Fourier transforms $\hat{u}_{m}$ above directly we will
use the following formula which is easily verified for the general case
$\Omega \subset \mathbbm{R}^{n}$:
\[ F_{N} ( k ) = \frac{1}{( 2 \pi )^{n}}   \sum_{m=1}^{N}   
\frac{|\int_{\Omega}   \tmop{div} ( e^{- i  \langle k,x \rangle}  u_{m} ( x ) ) 
   \, dx |^{2}}{ (|k|^{2} \um \lambda_{m}^{2})} .
\]
If $\Omega$ is sufficiently smooth then the numerators above are boundary
integrals which in our case $\Omega = [ 0, \pi ]$ are easily read off,
yielding ($\lambda_{m} =m^{2}$):
\[ F_{N} ( k ) =  \frac{2}{\pi^{2}} \sum_{m=1}^{N} \frac{m^{2}  ( 1- ( -1
   )^{m} \cos   ( \pi  k ) )}{( k^{2} -m^{2} )^{2}} . \]

\begin{figure}[h]
  \centering
  %\resizebox{502px}{334px}
  {\includegraphics{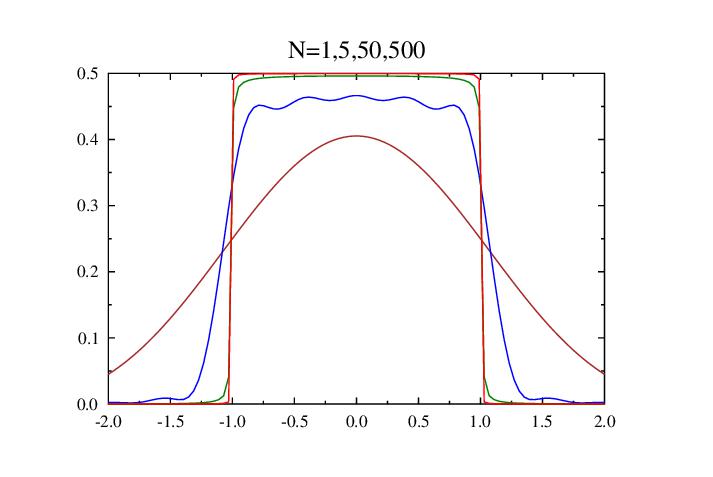}}
  \caption{}
\end{figure}

The figure shows the graphs of $F_{N} ( N k )$ for the values
$N=1,5,50,500.$ Thus one might ask

\begin{question}
  The diagram for $F_{N} ( N k )$, N$\in [ 1,5,50,500 ]$ suggests that \
  \[ \lim_{N \rightarrow \infty} F_{N} ( N k ) = \left\{\begin{array}{l}
       \frac{1}{2}   \ldots   \tmop{for}   | k | <1\\
       0  \ldots   \tmop{for}   | k | >1
     \end{array}\right. \]
  Does this hold for general dimensions and domains when suitably scaled? 
\end{question}

Indeed, for $| k | >1$, we have
\[ F_{N} ( N k ) = \frac{2}{\pi^{2}} \sum_{m=1}^{N} \frac{m^{2}  ( 1- ( -1
   )^{m} \cos   ( \pi  N k ) )}{( N^{2}  k^{2} -m^{2} )^{2}} \leqslant
   \frac{2}{\pi^{2}} \sum_{m=1}^{N} \frac{2 m^{2} }{N^{4} \left( k^{2} -
   \frac{m^{2}}{N^{2}} \right)^{2}} \leqslant \frac{4}{N^{4} \pi^{2} ( k^{2}
   -1 )^{2}} \sum_{m=1}^{N} m^{2} \]
\[ =  \frac{4N ( 2N^{2} +3N+1 )}{N^{4} \pi^{2} ( k^{2} -1 )^{2}}   \rightarrow
   0  \nocomma   \hspace{1.2em} \tmop{as}   \hspace{1.2em} N \rightarrow
   \infty . \]
For $k=0$ we get
\[ F_{N} ( N \cdot 0 ) = \frac{2}{\pi^{2}} \sum_{m=1}^{N} \frac{1- ( -1
   )^{m}}{m^{2}} , \]
therefore
\[ \lim_{N \rightarrow \infty}  F_{N} ( 0 ) =F_{\infty} ( 0 ) =
   \frac{2}{\pi^{2}} \left( \sum_{m=1}^{\infty} \frac{1}{m^{2}} -
   \sum_{m=1}^{\infty} \frac{( -1 )^{m}}{m^{2}} \right) = \frac{2}{\pi^{2}}
   \left( \frac{\pi^{2}}{6} + \frac{\pi^{2}}{12} \right) = \frac{1}{2} \]
as expected $\left( \frac{| \Omega |}{2 \pi} = \frac{\pi}{2 \pi} = \frac{1}{2}
\right) .$ Since $\int_{[ -1,1 ]} F_{N} ( N k )  d k=1$ and $0 \leqslant F
\leqslant \frac{1}{2}$ $\Rightarrow   \tmop{actually}  F_{N} ( N k ) =
\frac{1}{2}$ for $| k | <1$, at least almost everywhere.

To conclude the example we remark that for the cube $[ 0, \pi ]^{n} \subset
\mathbbm{R}^{n}$ we \ get the same conclusion when we consider the expression
\[ F_{N} ( N^{1/n}  k ) = \left( \frac{2}{\pi^{2}} \right)^{n} \sum_{m_{1}
   =1}^{M} \ldots \sum_{m_{n} =1}^{M} \prod_{j=1}^{n} \frac{m_{j}^{_{} 2}  (
   1- ( -1 )^{m_{j}} \cos   ( \pi  M k ) )}{( M^{2}  k^{2} -m_{j}^{2} )^{2}} ,
\]
where $N=M^{n} .$
\[  \]
\begin{remark}
  Repeating the calculations for $\left\{ e_{n} ( x ) = \frac{1}{\sqrt{2 \pi}}
  e^{i n x} \right\}_{n \in \mathbbm{Z}}$ on $[ 0,2 \pi ]$ we get for the
  Fourier transforms and the radius:
  \[ | \hat{e}_{n} ( k ) |^{2} = \frac{1}{2 \pi^{2}}   \frac{1-  \cos ( 2 \pi 
     k )}{( n-k )^{2}} , \hspace{0.8em} | \Omega | =2 \pi   \Rightarrow  
     \frac{| \Omega |}{( 2 \pi )} =1  \Rightarrow   \kappa_{F} ( \Omega ) =1.
  \]
  Consequently
  \[ F_{N} ( N k ) = \sum_{n=-N}^{N} | \hat{e}_{n} ( N k ) |^{2} =
     \sum_{n=-N}^{N}   \frac{1}{2 \pi^{2}}   \frac{2 \sin^{2} ( N \pi k )}{(
     n-N k )^{2}} = \frac{1}{\pi^{2}} \sum_{n=-N}^{N}   \frac{\sin^{2} ( N \pi
     k )}{( n-N k )^{2}} , \]
  and
  \[ \sum_{n=-N}^{N} | \hat{e}_{n} ( N k ) |^{2} \rightarrow \theta ( 1- | k |
     ) , \]
  which shows that the boundary conditions are not really relevant for the
  convergence.
  
  Taking the Haar system, $x \in [ 0,1 ] ,f_{0} ( x ) =1.$
  \[ f_{j,n} ( x ) =2^{\tfrac{n-1}{2}}   \chi_{\left[ \frac{2j-2}{2^{n}} ,
     \frac{2 j-1}{2^{n}} \right]} ( x ) -2^{\tfrac{n-1}{2}}   \chi_{\left[
     \frac{2j-1}{2^{n}} , \frac{2 j}{2^{n}} \right]} ( x ) \]
  $j=1 \ldots \nosymbol 2^{n-1}$, $n=1,2,3, \ldots$, we obtain
  \[ {\color{black} | \hat{f}_{j,n} ( k ) |^{2} = 2^{n+2}   \frac{  \sin^{4}  
     \left( \tfrac{k}{2^{n+1}} \right)}{  \pi  k^{2}} ,} \]
  which shows that not every Fourier image of a basis in $L^{2} ( \Omega )$,
  although in $P W_{\Omega} ( \mathbbm{R} )$, gives raise to the observed
  limit behaviour. 
\end{remark}

\subsection{Pointwise Convergence and Scheffe's Theorem}

Functions in $P W_{\Omega} ( \mathbbm{R}^{n} )$ are uniformly bounded and
vanish at infinity. The following simple thoughts show that this can be
sufficient under certain conditions to get convergence to a Heaviside
function.

Let $f_{n} :\mathbbm{R}^{d} \rightarrow [ 0,1 ]$ be a sequence of functions
such that
\begin{enumeratealpha}
  \item $\lim_{n \rightarrow \infty}  f_{n} ( x ) =1.$
  
  \item $\lim_{| x | \rightarrow \infty}  f_{n} ( x ) =0$.
\end{enumeratealpha}
Let $\{ a_{n} \} \subset \mathbbm{R} $ be a null sequence. From (b) we
conclude that there exists a sequence $\{ r_{n} \}$ such that
\[ f_{n} ( x ) \leqslant a_{n} \nocomma , \forall | x | \geqslant r_{n} . \]
Therefore we have
\[ f_{n} ( \lambda_{n} x ) \leqslant a_{n} , \forall   | x | \geqslant
   \frac{r_{n}}{\lambda_{n}} . \]
Thus, if for another sequence $\{ \lambda_{n} \}$ exists
\[ \limsup_{n \rightarrow \infty}   \frac{r_{n}}{\lambda_{n}} =x^{\star}   
\]
then
\[ \lim_{n \rightarrow \infty}  f_{n} ( \lambda_{n}  x ) =0, \forall | x |
   >x^{\star} . \]
On the other hand, suppose
\[ f_{n} ( x ) \geqslant 1-a_{n} , \forall | x | \leqslant \rho_{n} \]
then along the same lines:
\[ f_{n} ( \mu_{n}  x ) \geqslant 1-a_{n} , \forall   | x | \leqslant
   \frac{\rho_{n}}{\mu_{n}} . \]
\[ \liminf_{n \rightarrow \infty}   \frac{\rho_{n}}{\mu_{n}} =x^{}_{\star}  
   \Rightarrow   \lim_{n \rightarrow \infty}  f_{n} ( \mu_{n}  x ) =1, \forall
   | x | <x_{\star} . \]
Consequently
\[ \frac{r_{n}}{\lambda_{n}}   \sim \frac{\rho_{n}}{\mu_{n}}   \Rightarrow 
   x_{\star} =x^{\star}   \hspace{1.2em} i.e \nosymbol . \hspace{1.2em}
   \lim_{n \rightarrow \infty} \frac{r_{n}}{\lambda_{n}} = \lim_{n \rightarrow
   \infty}   \frac{\rho_{n}}{\mu_{n}} \]
That means, if we can show pointwise convergence then by Scheffe's theorem \cite{HS}
we also have convergence in measure (more precisely: the associated
densities).

The simple example $f_{n} ( x ) =1/ \left( 1+ \tfrac{x^{4}}{n^{4}} \right)
\nocomma , \; d=1$, however, gives
\[ \lim_{n \rightarrow \infty}  f_{n} \left( \tfrac{\pi n}{\sqrt{2}} x \right)
   =  \frac{1}{1+ \frac{\pi^{4} x^{4}}{4  }} , \]
in spite of $\int x^{2} f_{n} ( x ) \sim n^{3}$ is finite for all $n \in
\mathbbm{N}.$ Although the $f_{n}$ are not in $P W_{\Omega} ( \mathbbm{R} )$,
they are real analytic and have finite kinetic energy. Apparently, a universal
condition granting convergence to a ``Fermi ball'' cannot be stated easily.

\subsection{Weyl's Law and the Fermi Ball}

For the Dirichlet Laplacian $\left( \ref{eq10} \right)$ the asymptotics of the
eigenvalues $\lambda_{m}$ as $m \rightarrow \infty$ is well known
\begin{equation}
  \lambda_{m} = c_{n} ( \Omega )  m^{2/n}  +o ( 1 ) \label{Weyl}
\end{equation}
and so for the sum of the first $N$ eigenvalues
\[ \sum_{m=1}^{N} \lambda_{m} =c_{n} ( \Omega )   \frac{n}{n+2}  N^{1+
   \tfrac{2}{n}} +o ( 1 ) \]
where
\[ c_{n} ( \Omega ) = \frac{4 \pi   \Gamma \left( 1+ \tfrac{n}{2}
   \right)^{2/n}}{| \Omega |^{2/n}} \assign \kappa_{F}^{2} . \]
When we recall $\left( \ref{W2} \right)$ we see that the limit exists:

\[ \lim_{N \rightarrow \infty}   \sum_{j=1}^{N} \int_{\mathbbm{R}^{n}} | k
   |^{2}   | \hat{u}_{j} ( N^{1/n}  k ) |^{2}  d k=N^{1- \tfrac{2}{n}}  
   \sum_{j=1}^{N} \lambda_{j} =   \frac{n}{n+2}   \kappa_{F}^{2} . \]
\[  \]
The quantity $\kappa_{F} = \kappa_{F} ( \Omega )$ is sometimes called the
Fermi radius, i.e. more precisely the Fermi momentum ($\hbar =1$). It is the
radius of the ball $B_{\kappa_{F}}$ such that
\begin{equation}
  \frac{| \Omega |}{( 2 \pi )^{n}}   | B_{\kappa_{F}} | =1. \label{FB}
\end{equation}
Indeed, if we recall that the volume of the unit ball is given by
\[ \omega_{n} = \frac{\pi^{n/2}}{\Gamma ( 1+n/2 )} \]
then
\[ \frac{| \Omega |}{( 2 \pi )^{n}}   \omega_{n}   \kappa_{F}^{n} =1
   \Rightarrow   \kappa_{F} = \frac{2 \sqrt{\pi}   \Gamma ( 1+n/2 )^{1/n}}{|
   \Omega |^{1/n}} = \sqrt{c_{n} ( \Omega )} = \frac{2 \pi}{( \omega_{n} |
   \Omega | )^{1/n}} . \]
When we look at equation $\left( \ref{Wp} \right)$ then we might suspect that
for the fractional Laplacian (symbol $| k |^{p}$)
\[ \lim_{N \rightarrow \infty}   \sum_{j=1}^{N} \int_{\mathbbm{R}^{n}} | k
   |^{p}   | \hat{u}_{j} ( N^{1/n}  k ) |^{2}  d k=N^{1- \tfrac{p}{n}}  
   \sum_{j=1}^{N} \lambda_{j} =  \frac{n}{n+p}   \kappa_{F}^{p} \]
holds analogously:
\[ \sum_{j=1}^{N}   | \hat{u}_{j} ( N^{1/n}  k ) |^{2}   \xrightarrow[N
   \rightarrow \infty]{}   \frac{| \Omega |}{( 2 \pi )^{n}}  
   \chi_{B_{\kappa_{F}} ( k )} . \]
Actually, if we assume that
\[   \sum_{j=1}^{N} \int_{\mathbbm{R}^{n}} | k |^{p}   | \hat{u}_{j} (
   N^{1/n}  k ) |^{2}  d k \xrightarrow[N \rightarrow \infty]{} \frac{| \Omega
   |}{( 2 \pi )^{n}} \int | k |^{p}   \chi_{B_{\kappa_{F}} ( k )}  d k \]
\[  \]
then the last integral gives the correct asymptotic value.

\subsection{Mean Values}

\begin{definition}
  Let $u \in C ( \mathbbm{R}^{n} ) \nosymbol \nosymbol \nosymbol$.The
  spherical average $M_{u} ( x,r )$ over a sphere with radius $r$and center
  $x$ is defined as:
  \[ M_{u} ( x,r ) = \frac{1}{n  \omega_{n}} \int_{S^{n-1}} u ( x+r \theta ) 
     d \sigma ( \theta ) , \]
  where $\omega_{n} = \frac{2 \pi^{n/2}}{n  \Gamma \left( \frac{n}{2} \right)}
  $denotes the volume of the unit ball in $\mathbbm{R}^{n} .$
\end{definition}

Let $\Delta u+ \lambda u=0 $ in $\Omega .$ Then it is shown in \cite{CH}
\begin{equation}
  M_{u} ( x,r ) =u ( x )   \frac{\Gamma \left( \frac{n}{2} \right)
  J_{\frac{n-2}{2}} \left( \sqrt{\lambda} r \right)}{\left( \frac{r
  \sqrt{\lambda}}{2} \right)^{\frac{n-2}{2}}} =u ( x )  P_{n} \left(
  \sqrt{\lambda} r \right) . \label{MV1}
\end{equation}
for any sphere $\partial B_{r} ( x ) \subset \Omega .$ We write $P_{n}$ as
\[ P_{n} ( \xi ) = \frac{( 2 \pi )^{n/2}}{n  \omega_{n}}
   \frac{J_{\frac{n-2}{2}} ( \xi )}{\xi^{\frac{n-2}{2}}} . \]
It hold for example:
\[ P_{2} ( \xi ) =J_{0} ( \xi ) \]
and
\[ P_{3} ( \xi ) = \frac{\sin ( \xi )}{\xi} . \]
If $n$ is odd then $P_{n}$ may be expressed by derivatives of $P_{3} :$
\[ P_{2m+1} ( \xi ) = \frac{( -1 )^{m-1}  2^{2m-1}   \Gamma \left( m+
   \frac{1}{2} \right)}{\sqrt{\pi}} \frac{d^{m-1}}{d ( \xi^{2} )^{m-1}} \left(
   \frac{\sin ( \xi )}{\xi} \right) , \; m=1,2, \ldots \]
Moreover, let $\mu = \frac{n-2}{2}$, then
\[ \xi^{\mu} P_{2 \mu +2} ( \xi ) \sim J_{\mu} ( \xi ) , \]
using Bessel's differential equation:
\[ \xi^{2} ( \xi^{\mu} P_{2 \mu +2} ( \xi ) )'' + \xi ( \xi^{\mu} P_{2 \mu +2}
   ( \xi ) )' + ( \xi^{2} - \mu^{2} ) \xi^{\mu} P_{2 \mu +2} ( \xi ) =0 \]
thus ($\xi \neq 0$)
\[ P_{2 \mu +2} ( \xi )'' + \frac{2 \mu +1}{\xi} P_{2 \mu +2} ( \xi )' +P_{2
   \mu +2} ( \xi ) =0  \Longleftrightarrow \Delta_{n} P_{n} ( | x | ) +P_{n} (
   | x | ) =0. \]

\subsubsection{Radial Fourier Transform}

Since $u_{k} ( x ) =e^{- i k x}$ is a solution to $\Delta u ( x ) + | k |^{2} 
u ( x ) =0$ we must have by $\left( \ref{MV1} \right) :$
\[ M_{u_{k}} ( x,r ) =u_{k} ( x ) P_{n} ( | k |  r ) =e^{- i k x}  P_{n} ( | k
   | r ) . \]
Therefore $( \tmop{if}  f ( x ) =f ( | x | ) ) :$
\[ \hat{f} ( k ) = \frac{1}{( 2 \pi )^{n/2}} \int_{\mathbbm{R}^{n}} f ( x ) 
   e^{-i  \langle k,x \rangle}  d x= \frac{n \omega_{n}}{( 2 \pi )^{n/2}}
   \int_{0}^{\infty} f ( r )  M_{u_{k}} ( 0,r )  r^{n-1}  d r. \]
so that again
\begin{equation}
  {\color{black} {\color{black} \hat{f} ( \kappa ) = \frac{n \omega_{n}}{( 2
  \pi )^{n/2}} \int_{0}^{\infty} f ( r )  P_{n} ( \kappa  r )  r^{n-1}  d r.}
  \label{RFT1}}
\end{equation}
This may also be written as
\begin{equation}
  {  \hat{f} ( k ) = \hat{f} ( | k | ) =
  \frac{1}{( 2 \pi )^{n/2}}   \int_{\mathbbm{R}^{n}} f ( | x | )  P_{n} ( | k
  |   | x | )  d x.} \label{RFT2}
\end{equation}
It is obvious that the inverse transform has the same form.

\subsubsection{FT of the Characteristic Function of a Ball}

Let $f ( r ) = \left\{\begin{array}{l}
  a  \tmop{for}  r \leqslant \lambda\\
  0  \tmop{else} .
\end{array}\right.$, then
\[ {\color{black} \hat{f} ( \kappa ) = \frac{a n \omega_{n}}{( 2 \pi )^{n/2}}
   \int_{0}^{\lambda}  P_{n} ( \kappa  r )  r^{n-1}  d r {\color{black} =
   \frac{a n \omega_{n}}{\kappa^{n} ( 2 \pi )^{n/2}} \int_{0}^{\kappa  
   \lambda}  P_{n} ( \rho )   \rho^{n-1}  d  \rho .}} \]
\[ = \frac{a }{\kappa^{n} ( 2 \pi )^{n/2}} \int_{| x | < \kappa \lambda} P_{n}
   ( | x | )  d x= \frac{- a n  \omega_{n}   ( \kappa \lambda
   )^{n-1}}{\kappa^{n} ( 2 \pi )^{n/2}}   \left( \frac{d}{d s} P_{n} ( s )
   \right) |_{s= \kappa \lambda} \nobracket \]
where we have used that $\Delta P_{n} +P_{n} =0.$
\[ \frac{d}{d s} P_{n} ( s ) =- \frac{( 2 \pi )^{n/2}}{n  \omega_{n}}
   \frac{J_{\frac{n}{2}} ( s )}{s^{\frac{n-2}{2}}} \]
thus
\begin{equation}
  {\color{black}   \widehat{\chi_{}}_{B_{\lambda}} ( \kappa )} =
  {\color{black}   \left( \frac{\lambda}{\kappa} \right)^{n/2} J_{\frac{n}{2}}
  ( \kappa \lambda )} \label{CHB}
\end{equation}

\subsection{Kernels and Autocorrelation}

To each eigenfunction $u_{m}$ of $\left( \ref{eq10} \right)$ we associate the
function
\[ U_{m} ( x ) = \frac{1}{( 2 \pi )^{n/2}} \int_{\mathbbm{R}^{n}} u_{m} ( x+y
   )   \bar{u}_{m} ( y )  d y \]
which has its support in $\Omega \oplus ( - \Omega )$. The Fourier transform
of these functions are given by
\[ \hat{U}_{m} ( k ) = | \hat{u}_{m} ( k ) |^{2}  ,m \in \mathbbm{N}, \]
so that we can write
\[ F_{N} ( N^{1/n} k ) = \sum_{m=1}^{N} | \hat{u}_{m} ( N^{1/n}  k ) |^{2} =
   \sum_{m=1}^{N} \hat{U}_{m} ( N^{1/n}  k ) . \]
When we actually perform the Fourier transform of $U_{m}$ we obtain
\begin{equation}
  F_{N} ( N^{1/n} k ) = \frac{1}{N  ( 2 \pi )^{n/2}} \sum_{j=1}^{N}
  \int_{\mathbbm{R}^{n}} U_{j} \left( \tfrac{\xi}{N^{1/n}} \right)  e^{- i 
  \langle k, \xi \rangle}  d  \xi . \label{FTUM}
\end{equation}
Now let $Q_{N} ( x,y )$ denote the kernel
\[ Q_{N} ( x,y ) = \frac{1}{N} \sum_{m=1}^{N} u_{m} \left( x+
   \frac{y}{2N^{1/n}} \right)  u_{m} \left( x- \frac{y}{2N^{1/n}} \right) . \]
B. Schmidt has proven in \cite{BS}
\begin{equation}
  Q_{N} \xrightarrow[N \rightarrow \infty]{}  Q_{\star} = ( 2 \pi )^{-n}  
  \chi_{\Omega}   \otimes \hat{\chi}_{B_{\kappa_{F}}}   \label{BS}
\end{equation}
in $L^{2} ( \mathbbm{R}^{n} \times \mathbbm{R}^{n} )$ if $\Omega$ is such that
Weyl's Law is valid. Actually it is shown
\[ Q_{\star} ( x,y ) = \chi_{\Omega} ( x )   \frac{2^{n/2}   \Gamma \left( 1+
   \tfrac{n}{2} \right)  J_{\tfrac{n}{2}} ( \kappa_{F} | y | )}{| \Omega |   (
   \kappa_{F}   | y | )^{n/2}}  . \]
The equivalence to $\left( \ref{BS} \right)$ follows from $\left( \ref{CHB}
\right)$ and $\left( \ref{FB} \right)$.

Setting
\[  G_{N} ( x ) := \frac{1}{N} \sum_{j=1}^{N} U_{j} \left(
      \tfrac{x}{N^{1/n}} \right) = \frac{1}{N  ( 2 \pi )^{n/2}} \sum_{j=1}^{N}
      \int_{\Omega} u_{j} \left( \tfrac{x}{N^{1/n}} +y \right)  u_{j} ( y )  d
      y \]
   \[ = \frac{1}{N ( 2 \pi )^{n/2}} \sum_{j=1}^{N} \int_{\Omega +
      \tfrac{x}{N^{1/n}}} u_{j} \left( \xi + \tfrac{x}{2 N^{1/n}} \right) 
      u_{j} \left( \xi - \tfrac{x}{2 N^{1/n}} \right)  d  \xi \]
   \[ = \frac{1}{( 2 \pi )^{n/2}} \int_{\Omega + \tfrac{x}{N^{1/n}}} Q_{N} (
      \xi ,x )  d \xi \]
we can see that when comparing to $\left( \ref{FTUM} \right)$:
\[ \hat{G}_{N} ( k ) =F_{N} ( N^{1/n} k ) \]
and consequently by $\left( \ref{BS} \right)$ and Parseval's theorem we get as
$N \rightarrow \infty$ :
\begin{equation}
  G_{N} ( x ) \xrightarrow[L^{2}]{} \frac{\hat{\chi}_{B_{| k_{F} |} ( x )}}{(
  2 \pi )^{n/2}} | \Omega |   \Longleftrightarrow    F_{N} ( N^{1/n} k )
  \xrightarrow[L^{2}]{}   \frac{| \Omega |}{( 2 \pi )^{n}}  
  \chi_{B_{\kappa_{F}} ( k )} . \label{L2}
\end{equation}
In summary we can state that for the example problem $\left( \ref{eq10}
\right)$ the scaled partial sums of $| \hat{u}_{m} |^{2}$ convergence strongly
in $L^{2}$ to the ``Fermi ball''  $\frac{| \Omega |}{( 2 \pi )^{n}}  
\chi_{B_{\kappa_{F}} ( k )}$. This is far more than the expected convergence
in measure (vaguely as well as weakly). It seems that the methods of
\cite{BS} may succeed for other symbols as well.

\subsubsection{Rate of decay }

If we assume $\left( \ref{L2} \right)$ then
\[ \lim_{N \rightarrow \infty} \left\| \hat{G}_{N} - \frac{| \Omega |}{( 2 \pi
   )^{n}}   \chi_{B_{\kappa_{F}}} \right\|_{L^{2} ( \mathbbm{R}^{n} )}
   \rightarrow_{}  0. \]
This means
\[ I_{N} = \int_{\mathbbm{R}^{n}} | F_{N} ( N^{1/n} k ) - \frac{| \Omega
   |}{( 2 \pi )^{n}}   \chi_{B_{\kappa_{F}}} ( k ) |^{2}  d k 
   \xrightarrow[N \rightarrow \infty]{}  0. \]
When expanding while using $\left( \ref{FB} \right)$ we get
\[ I_{N} = \frac{1}{N} \int_{\mathbbm{R}^{n}} F_{N} ( k )^{2}  d k-2 \frac{|
   \Omega |}{( 2 \pi )^{n}}   \int_{| k | \leqslant \kappa_{F}} F_{N} (
   N^{1/n} k )  d k+ \frac{| \Omega |}{( 2 \pi )^{n}} . \]
Recall that the $F_{N} ( k )$ are uniformly bounded by $\frac{| \Omega |}{( 2
\pi )^{n}}$ $( \tmop{that}' s  \tmop{why}  F_{N} \in L^{1} \cap L^{2} )$. Thus
\[ I_{N} \leqslant 2  \frac{| \Omega |}{( 2 \pi )^{n}} \left[ 1- \int_{| k |
   \leqslant \kappa_{F}} F_{N} ( N^{1/n} k )  d k \right] . \]
Whence we see that the tricky bit for a direct prove is to show that
\begin{equation}
  \liminf_{N \rightarrow \infty} \int_{| k | \leqslant \kappa_{F}} F_{N} (
  N^{1/n} k )  d k=1. \label{hyp}
\end{equation}
Because
\[ \int_{\mathbbm{R}^{n}} F_{N} ( N^{1/n} k )  d k=1, \]
$\left( \ref{hyp} \right)  $means that the decay of $F_{N}$ must be strong
enough (but not too strong) to concentrate uniformly in $B_{\kappa_{F}}$. The
well known concentration compactness lemma reveals that we need ``tightness''
so that the ``momentum'' cannot run away to infinity.

\subsection{Polya's Conjecture}

To conclude this introductory setion we have a look at the connection of the
sequence $\{ F_{N} \}_{N \in N}$ to the famous conjecture of Polya: is $\left(
\ref{Weyl} \right)$ a lower bound to the eigenvalues $\lambda_{m} ( \Omega )$
of $\left( \ref{eq10} \right)  ?$ That is, does
\[ \lambda_{m} ( \Omega ) \geqslant  m^{2/n}   \kappa_{F}^{2} \]
hold for all $m \in \mathbbm{N}?$ Polya himself proved this for the case when
$\Omega$ is a ``tiling'' domain. This problem is (to our knowledge) still
unsolved and even the case of the disk in $\mathbbm{R}^{2}$ has not been
settled yet (here too, as far as we know). An alternative formulation is, for
example, using $\left( \ref{FB} \right)$:
\[ \frac{| \Omega |}{( 2 \pi )^{n}}   | B_{\sqrt{\lambda_{m}}} |  
   \geqslant  m ? \]
which, if true, would imply
\[ | B_{\sqrt{\lambda_{m}}} |   \geqslant  m  | B_{\kappa_{F}} | .
\]
Now remembering the fact that
\[ \lim_{N \rightarrow \infty} \sum_{j=1}^{N}   | \hat{u}_{j} ( k ) |^{2}  =
   \frac{| \Omega |}{( 2 \pi )^{n}} , \]
the following equality for each $m$ is easily established when integrating
over the ball $B_{\sqrt{\lambda_{m}}}$ :
\[ {\color{black} \frac{| \Omega |   | B_{\sqrt{\lambda_{m}}} |}{(
   2 \pi )^{n}} - \sum_{j=1}^{m} \| \hat{u}_{j} \|^{2}  =
   \sum_{j=m+1}^{\infty} \int_{B_{\sqrt{\lambda_{m}}}} | \hat{u}_{j} ( k )
   |^{2}  - \sum_{j=1}^{m} \int_{\mathbbm{R}^{n} \backslash \nobracket
   B_{\sqrt{\lambda_{m}}}} | \hat{u}_{j} ( k ) |^{2} .} \]
Thus if we could show that
\[ \sum_{j=1}^{m} \int_{\mathbbm{R}^{n} \backslash \nobracket
   B_{\sqrt{\lambda_{m}}}} | \hat{u}_{j} ( k ) |^{2}   \leqslant
   \sum_{j=m+1}^{\infty} \int_{B_{\sqrt{\lambda_{m}}}} | \hat{u}_{j} ( k )
   |^{2} \]
holds for all $m \in \mathbbm{N}$, then Polya was right. Using the information
of the kintetic terms we get analogously
\[ \frac{| \Omega |  | B_{\sqrt{\lambda_{m}}} |}{( 2 \pi )^{n}}  
   \frac{\lambda_{m}}{n+2} - \sum_{j=1}^{m} \lambda_{j} \| \hat{u}_{j} \|^{2} 
   =  \sum_{j=m+1}^{\infty} \int_{B_{\sqrt{\lambda_{m}}}} | k |^{2} |
   \hat{u}_{j} ( k ) |^{2} - \sum_{j=1}^{m} \int_{\mathbbm{R}^{n} \backslash
   \nobracket B_{\sqrt{\lambda_{m}}}} | k |^{2} | \hat{u}_{j} ( k ) |^{2} . \]
It is not to overlook that this represents quite a subtle balance problem that
needs far more precise growth information of the $\hat{u}_{j}$ than those
which are presently known (see e.g. \cite{WB} and ref. therein).

\tt{V 1.0, kp@scios.ch}

\end{document}